\documentclass[a4paper,fleqn,usenatbib]{mnras}

\usepackage{newtxtext,newtxmath}

\usepackage[T1]{fontenc}
\usepackage{ae,aecompl}

\usepackage{graphicx}	
\usepackage{amsmath}	
\usepackage{amssymb}	

\title[Dependence of pulsar death line on the equation of state]{Dependence of pulsar death line on the equation of state}

\author[X. Zhou et al.]{
Xia Zhou,$^{1,2}$\thanks{E-mail: zhouxia@xao.ac.cn (XZ)}
Hao Tong,$^{3}$
Cui Zhu,$^{1,4}$
Na Wang,$^{1,2}$
\\
$^{1}$Xinjiang Astronomical Observatories, Chinese Academy of Sciences, Urumqi 830011, China\\
$^{2}$Key Laboratory of Radio Astronomy, Chinese Academy of Sciences\\
$^{3}$School of Physics and Electronic Engineering, Guangzhou University, Guangzhou 510006, China\\
$^{4}$University of Chinese Academy of Sciences, 19A Yuquan Road, Beijing 100049, China\\
}

\date{Accepted XXX. Received YYY; in original form ZZZ}

\pubyear{2015}

\begin{document}
\label{firstpage}
\pagerange{\pageref{firstpage}--\pageref{lastpage}}
\maketitle

\begin{abstract}
Pulsar death line can be defined in $P-\dot{P}$ diagram. Traditionally, radio-loud pulsars are supposed to locate above the death line where is the radio-loud region. With the development of observational equipment, the observational properties of the neutron star are remarkably diverse. In $P-\dot{P}$ diagram, some of the special sources are radio-quiet but lie above the death line. From the definition of the pulsar death line, different equation of states for neutron star or strange star results in different death lines. We discuss the influence of the equation of state on the pulsar death line and the possible link between different neutron star groups. The results show that central compact objects would be the small mass of self-bound strange stars, and rotating radio transients might be old pulsars on the verge of death. We suggest that PSR J2144-3933 is likely to be a large mass pulsar, which would be larger than $2.0\rm{M_{\odot}}$. Multiple observational facts would help us to reveal the nature of pulsars.
\end{abstract}



\section{Introduction}           
\label{sect:intro}

\begin{figure}
\includegraphics[width=\columnwidth]{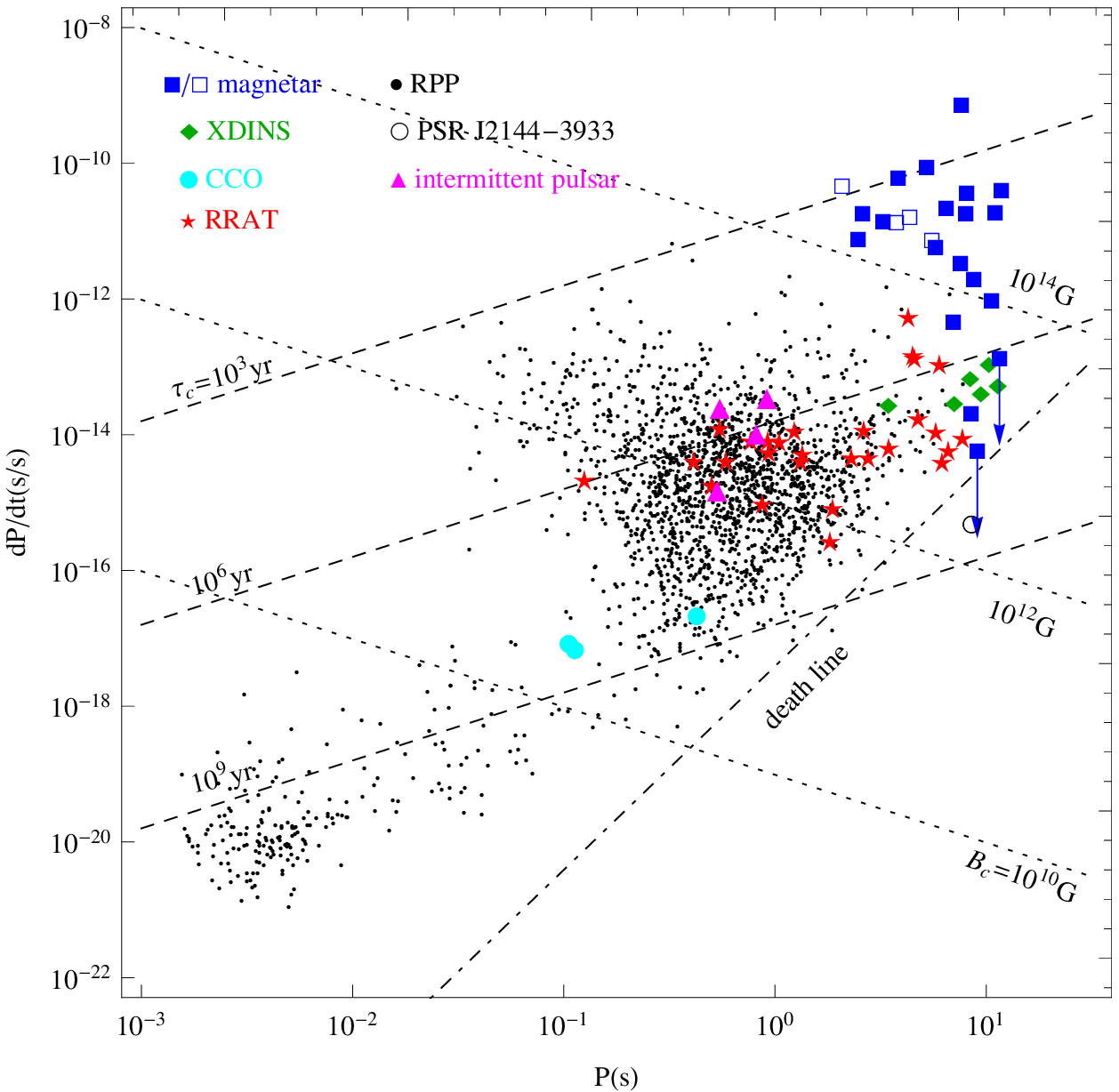}
\caption{The $P - \dot{P}$ diagram. Blue squares are magnetars (empty squares are radio-loud magnetars, from McGill magnetar catalog: \url{http://www.physics.mcgill.ca/pulsar/magnetar/main.html}). Green diamonds are X-ray dim isolated neutron stars (XDINSs, from \citealt{halpern2010}, \citealt{kaplan2011} and references therein), cyan circles are the central compact objects (CCOs, from \citealt{halpern2010,gotthelf2013}), red stars are Rotating Radio Transients (RRATs, from \url{http://astro.phys.wvu.edu/rratalog/}), magenta triangles are intermittent pulsars (From \citealt{lorimer2012} and references therein, \citealt{surnis2013}), black dots are rotation powered pulsars (including normal pulsars and millisecond pulsars, which dominate the pulsar population, from ATNF: \url{http://www.atnf.csiro.au/research/pulsar/psrcat/}), and the black open circle are PSR J2144-3933. Lines of characteristic magnetic field (dotted line), characteristic age (dashed line) and typical pulsar death line (dash-dot line) are also indicated. This figure is updated from work \citet{tong2014}.}
\label{fig1}
\end{figure}

Radio pulsars are unique laboratories for a wide range of physics and astrophysics, which were compellingly identified as ``classical'' neutron star (NS) not long after their discovery \citep{glendenning2000}. Its spin period $P$, and the period derivative $\dot{P}$, can be obtained at very high levels of precision through timing measurements \citep{manchester2005}. Then the pulsars could be placed on the $P-\dot{P}$ plane, which is a useful starting point for understanding the diverse flavors of pulsars. Figure~\ref{fig1} shows the $P-\dot{P}$ diagram for the known radio pulsars with black dots. 

It is generally realized that the radio emission from a pulsar originates from the production of electron-positron pairs in its magnetosphere \citep{ruderman1975, arons1979}. The magnetic field of a pulsar is rarely aligned with the spin axis, thus the observed pulses are produced by the lighthouse effect where the emission beam of the pulsar sweeps past our line of sight once per rotation. The pair-produced particles radiate energy, taking away the rotational energy. As the pulsar spins down and the unipolar potential drops, the accelerated particles achieve less energies and their subsequent curvature radiation or inverse Compton scattering photons are less energetic. Eventually, these electrons/positrons cannot be pair-produced. No further pairs are ejected into the plasma, and the pulsar stops radio emission. A radio pulsar dies \citep{ruderman1975, arons1979}. That means the potential drop across the accelerator, $\Delta V$, required for pair production exceeds the maximum one which can be achieved in the magnetosphere of a rotating pulsars, if the radio emission turned off. This maximum potential drop is written as \citep{ruderman1975}:
\begin{equation}
\label{eq1}
\Phi_{\rm{max}} \approx \frac{B_{\rm p} R^3 \Omega^{2}}{2c^2}
\end{equation}
where $R$ is the radius of a pulsar, $\Omega$ is the angular velocity of the stars and $c$ is the speed of light, $B_{\rm p}$ is the polar magnetic field strength at pulsar surface, conventionally determined by \citep{shapiro1983}
\begin{equation}
\label{eq2}
B_{\rm{p}}=\frac{1}{\sin \alpha}\left(\frac{3Ic^3 P\dot{P}}{2\pi^2 R^6}\right)^{1/2}
\end{equation}
where $I$ is the moment of inertia and $\alpha$ is the inclination angle. 
As a consequence, the so-called radio pulsar death line is defined by setting $\Delta V = \Phi_{\rm{max}} $, which is a line in the $P-\dot{P}$ diagram and separates the pulsars that can support pair production in their inner magnetosphere from those that cannot \citep{ruderman1975, zhang2000, zhang2003}. As shown in Equations~\ref{eq1} and \ref{eq2}, the maximum potential drop is also related to the moment of inertia and radius of NS, which is determined by the equation of state (EoS) of NS matter. In Figure~\ref{fig1}, the death line is plotted with the typical parameters of NSs ($M=1.4\rm{M_{\odot}}, R=10\rm{km}$) and constant potential drop $10^{12} \rm{V}$. In fact, it is difficult to define an exact death line in the $P-\dot{P}$ diagram, due to many uncertainties inherited to the death line. When considering the influence of the frame-dragging effect on the inner gap model, a coefficient can be added to right of the equal sign in Equation~\ref{eq2} \citep{zhang2000}. For different particle acceleration model above a pulsar polar cap, the potential drop $\Delta V$ are much more different. And the particle acceleration models have been undergoing significant revision in the former works \citep{chen1993,zhang1996, zhang1997,harding1998, gil2001, gil2002, wu2003}. In addition to this, there are also many other factors that would affect the position of pulsar death line, such as the configuration of surface magnetic field and EoS \citep{zhang2003}.

With the development of the observational equipment, the number of new radio pulsars with a wide range of parameters dramatically increased. From optical to $\gamma$-ray, there exist plenty of observational phenomena leading to identifications of other classes of NSs have been identified, such as magnetars, X-ray dim isolated neutron stars (XDINS), central compact objects (CCOs) and Rotating Radio Transients (RRATs) \citep{kaspi2010, kaspi2016}. Some of them have precise measurements of $P$ and $\dot{P}$, and we plot them in Figure~\ref{fig1} with different symbols and colors. 

Magnetars are assumed to be NSs that are powered by their strong magnetic field \citep{duncan1992}. Observationally, they have a rotational period in the range of $2-12 \rm{s}$ and their period derivative span a broader range from $10^{-14}$ to $10^{-10}$. This scatter of period derivative results in a scatter of the characteristic magnetic field of magnetars from less than $10^{13}\rm{G}$ to higher than $10^{15} \rm{G}$. They are young objects with characteristic
ages from $\sim10^ 3 \rm{yr}$ to more than $10^ 7 \rm{yr}$ \citep{tong2016,kaspi2017}. There are only four magnetars observed in radio band \citep{camilo2007, levin2012}. 

XDINSs are sub-classes of NSs because they are isolated neutron stars and with X-ray emission. The quasi-thermal X-ray emission of XDINSs are relatively low, great proximity, lack of radio counterpart, and relatively long periodicities ($P = 3-11 \rm{s}$). Timing observations of several objects have revealed that they are spinning down regularly, with typical inferred dipolar surface magnetic fields $\sim 1-3 \times 10^{13} \rm{G}$ and characteristic ages of $\sim 1-4 \rm{Myr}$ \citep{haberl2007, kaplan2009}. Up to now, no radio emission of XDINSs is detected. They are located in the upper right of the pulsar $P-\dot{P}$ diagram, between standard radio pulsars and magnetars. 

CCOs are neutron-star-like objects at the centers of supernova remnants. Common properties for CCOs are characterized by steady flux, predominantly surface thermal X-ray emission, lack of a surrounding pulsar wind nebula, and absence of detection at any other wavelength \citep{deluca2008, gotthelf2009, halpern2010}. Of the eight most secure CCOs, three are known to be neutron stars with spin periods of $0.105, 0.424$, and $0.112 \rm{s}$ \citep{gotthelf2013}. Spin-down rate has been detected for three CCOs, and they are placed in the middle of the $P-\dot{P}$ diagram. 

RRATs are a curious class of Galactic radio sources which characterized by strong radio bursts at repeatable dispersion measures, but not detectable using standard periodicity-search algorithms. Nevertheless, the observed pulses are inferred to occur at multiples with an underlying periodicity that is very radio-pulsar-like \citep{mcLaughlin2006, keane2011}. Their periods distribute over a wide range, from $0.7 \rm s$ to $6.7 \rm s$ \citep{keane2010}. RRATs occupy the region partly coinciding with the main radio pulsar cloud in the $P-\dot{P}$ diagram (Figure~\ref{fig1}), and partly extending towards longer spin periods and larger magnetic fields, but mainly fall into the end range of the conventional radio pulsars. Intermittent pulsars are having higher slow down rates in the on state (radio-loud) than in the off state (radio-quiet), and suggested to be part-time radio pulsars \citep{kramer2006}. Their nulling time scales are very long compared with that of the ordinary pulsars, radiation switching between the on state and the off state. Many models have been proposed for intermittent pulsars \citep{li2006,zhang2007, jones2012, li2014}, but it is still an open question as to what is accounts for the on/off transition and why the rotation slows down faster when the pulsar is on than when it is off. 
 
As discussed before, our knowledge about NSs increases a lot thanks to rapid development of observational technique during the past two decades. Radio pulsars are just one observational manifestation of NSs. Some other groups of the sources are lying above the death line but they are radio-quiet, which are shown in Figure~\ref{fig1}, such as XDINSs,  CCOs and some magnetars. Although none of the XDINSs and CCOs has been detected at radio frequencies, it does not necessarily mean that they have no radio emission. Some authors argued that XDINSs are actually radio pulsars viewed well off from the radio beam \citep{kaspi2016}. CCOs have no optical or radio counterpart or pulsar wind nebula, no radio emission are detected, beaming effect cannot explain this lack of radio emission, ultra low radio luminosities could. Besides beaming effect and ultra low radio luminosities, it seems going back to the basic concept of pulsar death line could help us to understand these phenomena from another perspective. 

The standard definition of a death line implicitly assumes that pair formation is a necessary condition for pulsar radio emission and that pulsars become radio-quiet after crossing the death lines during their evolution from left to right in the $P-\dot{P}$ diagram. Obviously, this basic condition is not sufficient and functioning of radio pulsars may imply far more complex physical conditions. However, numerous theoretical attempts to produce satisfactory death lines implying even the basic necessary condition meets certain challenges \citep{harding2001,harding2002,usov2002}. Most of previous works are mainly focussed on the configurations of surface magnetic field and the models for pair acceleration. Moreover, the potential drop $\Delta V$, across an accelerator gap, is model dependent. Different particle acceleration model will change $\Delta V$ considerably and alter the death lines. The potential drop $\Delta V$ just a few $10^{12} \rm{V}$ \citep{ruderman1975}, or has a weak dependence on $\Omega$ \citep{xu2001}. Meanwhile, a potential drop around $10^{12} \rm{V}$ is also a widely accepted result from the pulsar death line criterion \citep{zhang2000}. In order to minimize the underlying model assumptions, we assume a constant potential drop $\Delta V=10^{12} \rm{V}$ in the polar cap accelerating region. As shown in Equation~\ref{eq1} and~\ref{eq2}, the pulsar death line is also related to the moment of inertia of NS, which is determined by the EoS. By setting the constant potential drop $\Delta V$, we only need to consider the dependence of pulsar death line on the EoS. Since the deconfinement transition from hadron matter into quark matter is possible at high density, one logically proposes that some pulsars could be strange stars(SSs). But it is still an open question whether these stars are NSs or SSs\citep{ozel2016}. The most convincing proof for a SS would come from a direct indication of the EoS, and we believe that pulsar death line might provided us the answer. 

In this work, we are going to discuss the properties of death lines affect by EoSs without considering the magnetosphere and particle acceleration model. Observational constraints on EoSs are also considered. Furthermore, the possible connections between different kinds of NSs will also be discussed. In Section 2, we will introduce the significance of EoSs in astrophysics and the different pulsar death lines with the impacts of EoSs. Conclusion and discussion will be provided in Section 3.

\section{EoS and pulsar death line}
\label{sect:eos}
\par
The nature of matter at the inside of NS is one of the major unsolved problems in modern science, and this makes NS unparalleled laboratories for nuclear physics and quantum
chromodynamics under extreme conditions \citep{glendenning2000}. Consequently, measurements of the properties of NSs provide constraints on the physics of cold dense matter that complements the constraints obtained from nuclear physics experiments \citep{watts2016, ozel2016, lattimer2017}. 

The most fundamental property of dense matter is the EoS, the pressure-density-temperature relation of bulk matter. Together with the Tolman-Oppenheimer-Volkoff (TOV) equations, the EoS of NS matter determines the macroscopic properties of the stars and their masses and radii. In fact, solutions provide a unique map between the microscopic pressure-density relation($p-\epsilon$ or $p-\rho$) and the macroscopic mass-radius ($M-R$) relation of the stars \citep{lindblom1992}.
This unique mapping can be used to infer the EoS from astrophysical measurements of their masses and radii \citep{watts2016}. The cleanest constraints on the EoS to date are derived from radio pulsar timing, where the mass of NSs in compact binaries can be measured very precisely using relativistic effects \citep{lorimer2008}. Finding the maximum mass of NSs provide additional direct constraints on the EoS. At a minimum, the EoSs that have maximum masses($\approx 2\rm{M_{\odot}}$, up to now) that fall below the most massive NS can be ruled out \citep{demorest2010, ozel2010, antoniadis2013}. 

\begin{figure}
\includegraphics[width=\columnwidth]{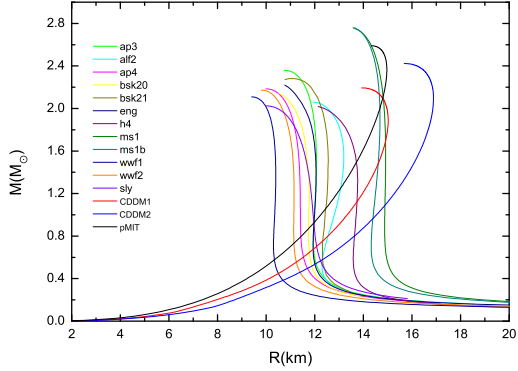}
\caption{The mass-radius curves corresponding to different EoSs which maximum mass are larger than $2.0\rm{M_{\odot}}$. The corresponding EoSs are get from \citet{ozel2016}(\url{http://xtreme.as.arizona.edu/NeutronStars/}), \citet{li2016} and \citet{bhattacharyya2016}.}
\label{fig2}
\end{figure}

Generally, two categories of EoS were widely discussed, which can produce gravity-bound NS and self-bound SS \citep{glendenning2000, haensel2007}. As shown in Figure~\ref{fig2}, the EoSs of self-bound SSs predicted $M \propto R^3$ for low-mass pulsars, meanwhile the radius is smaller. Moreover, the minimum mass of self-bound SSs can reach as low as planet mass \citep{xu2003, horvath2012}, while the low-limit mass of gravity-bound NSs is about $0.1\rm{M_{\odot}}$ \citep{akmal1997, glendenning2000}. On the other hand, searching for very low mass pulsars is also an attractive method, because the EoS of self-bound SSs predicted distinct radii at low mass compared with ones by the EoSs of gravity-bound SSs \citep{li2015, wang2017}. Thus, theoretical EoSs could be effectively tested from the accurate measurement of the radii for low-mass pulsars. In Figure~\ref{fig2}, we only showed the typical EoSs which have maximum mass larger than $2.0\rm{M_{\odot}}$. The typical nuclear matter EoSs are obtained from \citet{ozel2016}(\url{http://xtreme.as.arizona.edu/NeutronStars/}). 
The EoSs of SSs are get from \citet{li2016}(for CDDM1, CDDM2) and \citet{bhattacharyya2016}(for pMIT). 

\begin{figure}
\includegraphics[width=\columnwidth]{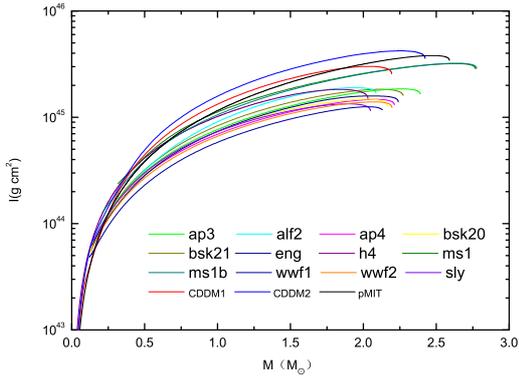}
\caption{The moment of inertia - mass curves corresponding to Figure~\ref{fig2}.}
\label{fig3}
\end{figure}

The pulsar death line is related to the NS moment of inertia $I$ and radius $R$ which is illustrated in Equation~\ref{eq1} and \ref{eq2}. With the equation $I=\frac{8\pi}{3}\int_{0}^{R}r^4\epsilon dr$ \citep{glendenning2000}, we calculated the NS moment of inertia by employing different EoSs, as shown in Figure~\ref{fig2}. The results are presented in Figure~\ref{fig3}. There is no obvious difference between the moment of inertia of NSs and SSs for both low and high mass. The distinction of moment of inertia between the large mass and small mass is very obvious.

As showed in Table~\ref{table1}, we calculate different radii and moment of inertias of the given mass of NSs. We specified the mass of NS, as $0.2\rm{M_{\odot}}, 0.5\rm{M_{\odot}}, 1.0\rm{M_{\odot}}, 1.4\rm{M_{\odot}}$ and $2.0\rm{M_{\odot}}$. The corresponding radii and moment of inertias are obtained by employing EoSs. The smallest mass of NSs with typical nuclear matter of EoSs named bsk21, wwf1, and h4 \citep{ozel2016}, sustained is $0.2\rm{M_{\odot}}$.

For the SSs, the radius increases with increasing mass. The smallest mass can be determined by the radius measurements. While measuring masses using radio pulsars are got precise results, measuring the radius is much more challenging and, with the methods currently in use, more model dependent. Current efforts to constrain the full $M-R$ relation focus primarily on spectroscopic measurements of the surface emission from accreting NSs in quiescence, and when they exhibit thermonuclear (type I) x-ray bursts due to unstable nuclear burning in accreted surface layers \citep{ozel2016}. Most CCOs have thermal-like X-ray emission, with blackbody temperatures in the range between $0.2-0.5 \rm{keV}$, no evidence for additional non-thermal components and many of them have relatively good distance estimates, which is important for measuring the radius \citep{gotthelf2013}. Better fit of the observation of CCOs in
the supernova remnant Cassiopeia A show the effective temperature
$kT ^{\infty}_{eff} \approx 0.2 \rm{keV}$, which require a small radius, $R = 4-5.5 \rm{km}$ \citep{pavlov2009}. XDINSs are characterized by thermal spectra with a broad spectral feature in most of them, which also could be used to measure the radius. The fitted radiation radius of RX J0806.4¨C4123 is about $3.5 \rm{km}$, and it is suggested to be self-bound SS \citep{wang2017}. Assuming the thermal emission comes from the whole surface, we could set the smallest radius of the SSs at $4\rm{km}$ in this work. The corresponding masses and moment of inertias are obtained by employing EoS named pMIT \citep{bhattacharyya2016}, CDDM1 and CDDM2 \citep{li2016}, which are shown in Table~\ref{table2}. For comparison, the mass of SSs are set as $0.03164\rm{M_{\odot}}, 0.02234\rm{M_{\odot}}, 0.01829 \rm{M_{\odot}}$, respectively (which is calculated from EoSs with $R=4\rm{km}$), $0.2\rm{M_{\odot}}, 0.5\rm{M_{\odot}}, 1.0\rm{M_{\odot}}, 1.4\rm{M_{\odot}}, 2.0\rm{M_{\odot}}$.

\begin{table}
\caption{The radius and moment of inertia for the specific mass of NSs with EoSs named bsk21, wwf1 and h4. The EoSs are got from \citet{ozel2016}(\url{http://xtreme.as.arizona.edu/NeutronStars/}).}
\label{table1}	
\centering 
\begin{tabular}{cccc}
	\hline\hline
name&M($M_{\odot}$) &R ($\rm{km}$)&I($g~cm^{2}$) \\
	\hline
	&0.2&15.66 &$9.288\times 10^{43}$ \\
	&0.5&12.43&$3.289\times 10^{44}$\\
bsk1&1.0&12.46&$8.411\times 10^{44}$\\
	&1.4&12.58&$1.280\times 10^{45}$\\
	&2.0& 12.28&$1.803\times 10^{45}$\\
	\hline
	&0.2&13.32 &$7.380\times 10^{43}$ \\
    &0.5&10.41&$2.338\times 10^{44}$\\
wwf1&1.0&10.34&$5.872\times 10^{44}$\\
   &1.4&10.39&$8.918\times 10^{45}$\\
   &2.0& 10.06&$1.255\times 10^{45}$\\
   \hline
 	&0.2&15.51 &$1.161\times 10^{44}$ \\
 &0.5&13.61&$4.142\times 10^{44}$\\
 h4&1.0&13.71&$1.017\times 10^{45}$\\
 &1.4&13.75&$1.509\times 10^{45}$\\
 &2.0& 12.38&$1.698\times 10^{45}$\\  
	\hline\hline
\end{tabular}
\end{table}%

\begin{table}
\caption{The radius and moment of inertia for the specific mass of SSs with EoSs named pMIT, CDDM1 and CDDM2. The EoS named pMIT are recovered from \citet{bhattacharyya2016},  EoSs named CDDM1 and CDDM2 are get from \citet{li2016}. The smallest mass is got with the best fit small radius of CCO. }
\label{table2}
\centering 
\begin{tabular}{cccc}
\hline\hline
name&M ($\rm{M_{\odot}}$) & R ($\rm{km}$) & I ($\rm{g~cm^2}$)  \\ 
\hline
&0.03164 &4.0     &$4.072\times 10^{42}$\\
&0.2         &7.358 &$8.549\times 10^{43}$ \\
pMIT&0.5         &9.888 &$3.825\times 10^{44}$\\
&1.0         &12.25 &$1.356\times 10^{45}$\\
&1.4         &13.48 &$1.931\times 10^{45}$\\
&2.0         &14.68 &$3.164\times 10^{45}$\\
\hline
&0.02234 &4.0     &$3.014\times 10^{42}$\\
&0.2         &7.981 &$1.176\times 10^{44}$ \\
CDDM1&0.5         &10.93 &$4.645\times 10^{44}$\\
&1.0         &13.34 &$1.335\times 10^{45}$\\
&1.4         &14.45 &$2.124\times 10^{45}$\\
&2.0         &14.97 &$3.002\times 10^{45}$\\
\hline
&0.01829 &4.0     &$2.644\times 10^{42}$\\
&0.2         &8.726 &$1.399\times 10^{44}$ \\
CDDM2&0.5         &11.84 &$5.579\times 10^{44}$\\
&1.0         &14.56 &$1.604\times 10^{45}$\\
&1.4         &15.87&$2.596\times 10^{45}$\\
&2.0         &16.86 &$3.954\times 10^{45}$\\
\hline\hline
\end{tabular}
\end{table}%

\begin{figure}
\includegraphics[width=\columnwidth]{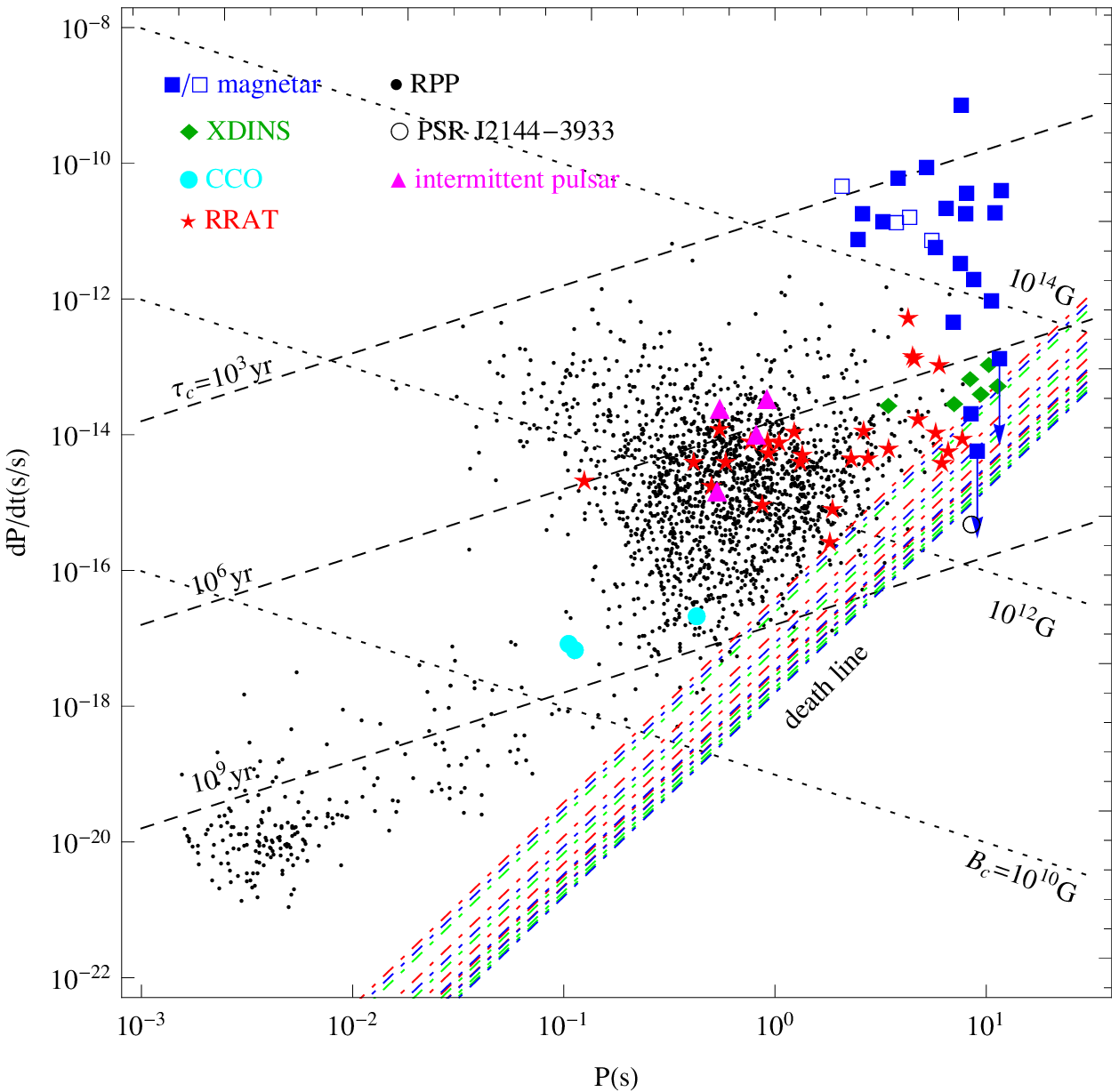}
\caption{Pulsar death lines in the $P - \dot{P}$ diagram for different neutron star masses and EoSs are shown in dash-dot red lines(bsk21), dash-dot green lines(wwf1) and dash-dot blue lines(h4). From top to bottom groups are for $0.2\rm{M_{\odot}}, 0.5\rm{M_{\odot}}, 1.0\rm{M_{\odot}}, 1.4\rm{M_{\odot}}$ and $2.0\rm{M_{\odot}}$. Also shown are sources which are the same as the caption of Figure~\ref{fig1} mentioned. The black open circle is PSR J2144-3933.}
\label{fig4}
\end{figure}

\begin{figure}
\includegraphics[width=\columnwidth]{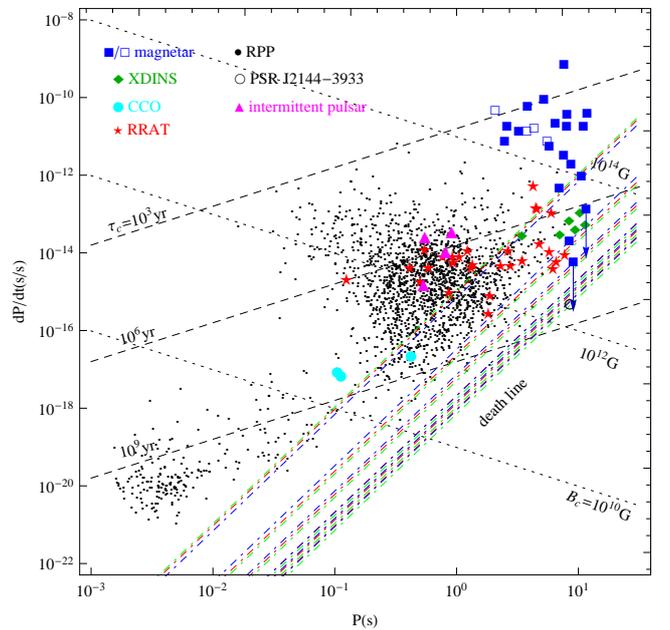}
\caption{Pulsar death lines in the $P - \dot{P}$ diagram for different EoSs and mass of strange stars with the dash-dot red lines(pMIT), dash-dot green lines(CDDM1) and dash-dot blue lines(CDDM2). From top to bottom groups are for $M(R=4\rm{km}), 0.2\rm{M_{\odot}}, 0.5\rm{M_{\odot}}, 1.0\rm{M_{\odot}}, 1.4\rm{M_{\odot}}$ and $2.0\rm{M_{\odot}}$. Also shown are sources which are the same as the caption of Figure~\ref{fig1} mentioned. The black open circle is PSR J2144-3933.}
\label{fig5}
\end{figure}

In Figure~\ref{fig4} and Figure~\ref{fig5}, we plot the pulsar death lines for different mass of NSs and SSs with constant potential drop ($\Delta V=10^{12} \rm{V}$) and the maximum one defined in Equation~\ref{eq1}. From top to bottom, the mass of the corresponding pulsar death line increases. For the same mass of NSs or SSs, death lines with different EoSs shows no obvious diversity with each others. The region of death lines of SS spreads wider than NS. The death lines of NSs and SSs do not have obvious difference while the mass is larger than $1.0\rm{M_{\odot}}$. Meanwhile, most of the sources are above the death lines. In Figure~\ref{fig4}, the long period ($8.5 \rm {s}$) pulsar PSR J2144-3933 is located beyond the death line, but closest to the death lines for $2.0\rm{M_{\odot}}$. PSR J2144-3933 just crosses the death lines with $2.0\rm{M_{\odot}}$ in Figure~\ref{fig5}. The obvious distinctions of the pulsar death line are for the smallest mass of NS and SS. In the following of this work, we focus on the pulsar death line with the smallest mass of NSs and SSs.

In Figure~\ref{fig4}, most of magnetars and XDINSs are above the death lines, and some of them are close to the death lines. In Figure~\ref{fig5}, all of the XDINSs are below or across the death lines, which agrees with the observational facts that XDINSs do not have detect radio emission until now. All of the radio-loud magnetars are above the death lines. And some of the radio-quiet magnetars are below the death lines. For the CCOs, in the NS case(Figure~\ref{fig4}), all of them are above the death lines. And in the case of SS (Figure~\ref{fig5}), they are near or below the death lines. RRATs are located around the normal pulsars, dispersed on the both side of the death lines. All of the intermittent pulsars are above the death lines for both NSs and SSs. 

\section{Discussion and conclusion}

Pulsars could radiate radio and high energy photons. There must be certain kinds of particle acceleration and radiation mechanisms in the pulsar magnetosphere. Previous studies showed that different polar gap models and geometry of the magnetic field will result in different pulsar death lines \citep{harding2002, zhang2000, wu2003}. Moreover, the potential drop across the accelerator and maximum potential drop are relying on the moment of inertia and radius of the stars \citep{xu2001}, to some degree. The M-R relations of the NS and SS are different, especially at low mass. EoSs affect the location of the pulsar death line in the $P-\dot{P}$ diagram. The results show that different EoSs results in different pulsar death lines. 

In Figure~\ref{fig2} and Figure~\ref{fig3}, we plotted the mass-radius and moment of inertia-mass curves with different EoSs. NSs and SSs differentiated from each other at the low mass region, due to the radius of self-bound SS is increasing with mass but gravity bound NS do not. Moreover, Figure~\ref{fig4} and Figure~\ref{fig5} show that the death lines of SS spread larger areas than that of NS. Particularly, the location of pulsar death lines with the smallest mass of SS is much higher than others. Comparing with Figure~\ref{fig4}, the distribution of sources, such as magnetars, XDINSs, CCOs, RRATs, is much more different for the death line with the smallest mass of SS. 

Magnetars have X-ray luminosity as high as $10^{35}~\rm{erg~s^{-1}}$, and their strong surface dipole field may lie in the range of $10^{13}-10^{15} \rm{G}$. And four of the magnetars are detected in the radio band. XDINSs are nearby sources with surface dipole field around $10^{13} \rm{G}$, very faint X-ray emission and without radio activities. In \citet{vigano2013}, the authors studied magneto-thermal evolution of isolated neutron stars. In this model, magnetars evolve and their magnetic fields dissipate, their observational properties (both timing and luminosities) appear compatible with those of the XDINSs. They suggested that XDINSs are supposed to be dead magnetars. In addition to these, two magnetars may have passive fallback disks\citep{wang2006,kaplan2009a}, and \citet{ertan2014} interpret that XDINSs have gone through a past accretion phase with spin-down and
emerged in their current state. In Figure~\ref{fig5}, we could find that the radio-loud magnetars are above the death line. Meanwhile, all of the XDINSs and some radio-quiet magnetars are below the death line. There might be some links between XDINSs and magnetars. We could suggest that XDINSs might be the result of magnetars evolution, even could be dead magnetars. From the study of pulsar death lines, the results here are in line with these previous speculations. Furthermore, well fitted spectra would provide the radii of the sources. The fitted radiation radius of RX J0806.4¨C4123 is about $3.5 \rm{km}$, and it could be a low-mass SS candidate \citep{wang2017}.

CCOs are known as well-fitted surface thermal X-ray emission and without emissions in the radio band. Due to its particular properties of thermal X-ray emission, CCOs are useful targets for measuring the temperatures and radii of the pulsars. The apparent blackbody radii of CCOs are only about a few kilometers, which is significantly smaller than the $10-15 \rm{km}$ canonical NS radii. The better fitted radius of CCO we used in this work are $R = 4-5.5 \rm{km}$. Actually, some of CCOs with well fitted blackbody radii are much smaller than we used here (data can be found from \citet{vigano2013} and Table in \citet{potekhin2015}, as well as from \url{www.neutronstarcooling.info}). In Figure~\ref{fig5}, the CCOs are near to or below the death line for smaller mass of self-bound SS. The CCOs supposed to be low mass SSs. Moreover, it is suggested that the radio-quiet CCO, 1E 1207.4-5209, could be a low mass bare strange star with a radius of a few kilometers according to its peculiar timing behavior \citep{xu2005}. If the smallest mass or radius of pulsars are detected in future, we also could suggest that they should be SSs. Well determined fitting model and sufficient observational data of thermal X-ray emissions would allow us to learn more about these thermal emitting sources.

RRATs generally have longer periods than those of the normal pulsar population, and several have high magnetic fields, similar to those other neutron star populations like the X-ray bright magnetars. However, some of the RRATs have spin-down properties very similar to those of normal pulsars, making it difficult to determine the cause of their unusual emission and possible evolutionary relationships between them and other classes of neutron stars. In the wind braking model \citep{kou2015, tong2017}, as the pulsar tends to death, the radio emission will not stop immediately, the pulsar will be observed if the condition of electron pair production is met. Moreover, \citet{zhang2007} suggested that the RRATs as well as the nulling pulsar and intermittent pulsars are located slightly below the death line and become occasionally active only when the conditions for pair production and coherent emission are satisfied. In Figure~\ref{fig5}, RRATs are dispersed on the both side of the death line for small mass SS. We might take RRATs as neutron stars on the verge of death. 

PSR J2144-3933 is a radio-loud pulsar with long period of $8.5 \rm s$ \citep{young1999}, and located far beyond all conventional death lines. And it seems to challenge the emission models. Many works discussed this 
problem by considering different types of the acceleration models \citep{zhang2000, gil2001}. In Figure~\ref{fig4} and \ref{fig5}, this pulsar is closed to and just cross to the pulsar death lines for $2.0\rm{M_{\odot}}$. In the traditional radio emission model, we could suggested that PSR J2144-3933 may has a high mass, which would be larger than $2.0 M_{\odot}$. 

As discussed in Section 2, constraints on EoSs will give us more informations about the pulsar death line which would help us to understand the observational properties of various classes of neutron stars. The maximum and minimum mass of the NSs measured from observational data provides additional direct constraints on the EoS, and the moment of inertia is a powerful probe of its internal structure and its EoS \citep{lattimer2005, raithel2016}. The dense matter EoS is the main scientific objective for a number of different telescopes over the next decade, operating in very different wavebands, such as the Square Kilometer Array (SKA, \citealt{bourke2015}),  the Five hundred meter Aperture Spherical radio Telescope (FAST, \citealt{nan2011}), Neutron Star Interior Composition Explorer Mission (NICER,  \citealt{gendreau2012}), Enhanced X-ray Timing and Polarization mission (eXTP, \citealt{zhang2016}) etc. For pulsar death lines, if the moment of inertia of the pulsar is well measured, they will provide a new avenue into our understandings of the pulsar death line. Till now, the most precise measurement of the mass of compact objects is in the binary systems, and isolated objects are very difficult to determine their masses. In this work, we used the X-ray luminosity to estimate the radii of the compact objects. The emission area which could only give us a lower limit of the surface area of the star. Comparing current radius measurements with those obtained from other techniques will further reduce the uncertainties in the NS radii. In this work, the pulsar death line is discussed with multi-wavelength observational properties, such as radio, X-ray. We expect that multiple observational facts would combine with each other to help us know more about pulsars.
 
\section*{Acknowledgements}

We gratefully thank Dr. Li Ang for useful discussions and comments. We also grateful to Dr. Rai Yuen for a careful read the manuscript and made a number of valuable suggestions, and to anonymous reviewer for the thoughtful comments that helped significantly improve the clarity of this work. This work is supported by the National Natural Science Foundation of China (No. 11373006), the West Light Foundation of Chinese Academy of Sciences (No. ZD201302). 






\bsp	
\label{lastpage}
\end{document}